\documentclass[prb,twocolumn]{revtex4}
\usepackage{bm}
\usepackage{dcolumn}
\usepackage{graphicx}
\begin{document}

\title{Dipolar interactions, molecular flexibility, and flexoelectricity
in bent-core liquid crystals}

\author{Alastair Dewar}

\author{Philip J. Camp}

\email{philip.camp@ed.ac.uk}

\affiliation{School of Chemistry, The University of Edinburgh, West Mains
Road, Edinburgh EH9 3JJ, United Kingdom}

\date{\today}

\begin{abstract} 
The effects of dipolar interactions and molecular flexibility on the
structure and phase behavior of bent-core molecular fluids are studied
using Monte Carlo computer simulations. Some calculations of flexoelectric
coefficients are also reported. The rigid cores of the model molecules
consist of either five or seven soft spheres arranged in a `V' shape with
external bend angle $\gamma$. With purely repulsive sphere-sphere
interactions and $\gamma=0^\circ$ (linear molecules) the seven-sphere
model exhibits isotropic, uniaxial nematic, smectic-A, and tilted phases.
With $\gamma \geq 20^\circ$ the smectic-A phase disappears, while the
system with $\gamma \geq 40^\circ$ shows a direct tilted
smectic--isotropic fluid transition. The addition of electrostatic
interactions between transverse dipole moments on the apical spheres is
generally seen to reduce the degree of tilt in the smectic and solid
phases, destabilize the nematic and smectic-A phases of linear molecules,
and destabilize the tilted smectic-B phase of bent-core molecules. The
effects of adding three-segment flexible tails to the ends of five-sphere
bent-core molecules are examined using configurational-bias Monte Carlo
simulations. Only isotropic and smectic phases are observed. On the one
hand, molecular flexibility gives rise to pronounced fluctuations in the
smectic-layer structure, bringing the simulated system in better
correspondence with real materials; on the other hand, the smectic phase
shows almost no tilt.  Lastly, the flexoelectric coefficients of various
nematic phases -- with and without attractive sphere-sphere interactions
-- are presented. The results are encouraging, but the computational
effort required is a drawback associated with the use of fluctuation
relations.
\end{abstract}
\maketitle

\section{Introduction}
\label{sec:introduction}

There has been a surging interest in bent-core liquid crystals since their
discovery in 1996 \cite{Niori:1996/a}. Typically, these materials consist
of molecules comprising rigid, banana-shaped cores made up of a conjugated
system of linked aryl groups, and flexible alkyl or alkoxy tails attached
to each end. The molecules are usually achiral and possess electric dipole
moments parallel with the molecular $C_{2}$ axes. One of the most
intriguing properties of these compounds is that, in some cases, chiral
ferroelectric or antiferroelectric smectic phases can be observed
\cite{Niori:1996/a,Macdonald:1998/a,Jakli:1998/a}. The chirality arises
because the molecules tilt within the smectic layers; in chiral
(anti)ferroelectric phases all of the molecules tilt in the same sense
with respect to the layer polarization vector. It is not yet clear what is
responsible for this spontaneous symmetry-breaking process, although a
variety of explanations has been proposed. One popular explanation
involves the long-range dipole-dipole interaction
\cite{Madhusudana:2004/a}, while recent theoretical work has identified a
central role for dispersion interactions \cite{Emelyanenko:2004/a}. Other
possible explanations include entropic `free-volume' mechanisms
\cite{Lansac:2003/a} in which an antiferroelectric ordering of the
smectic-layer polarizations affords more room for layer fluctuations, and
mechanisms in which the molecules themselves spontaneously select chiral
molecular conformations \cite{Earl:2005/a}.

There is a growing simulation literature on bent-core liquid crystals. One
of the most simple bent-core molecular models is a dimer made up of two
hard spherocylinders \cite{PJC:1999/d,Lansac:2003/a}. This system exhibits
isotropic, nematic, smectic, and crystalline phases, but no tilted phases.  
The Gay-Berne dimer model has been studied extensively, with and without
molecular dipoles. In the works by Memmer \cite{Memmer:2000/a} and
Johnston {\it et al.} \cite{Johnston:2002/a,Johnston:2002/b}, isotropic,
nematic, tilted smectic, and helical phases were found, depending on the
molecular bend angle \cite{Memmer:2000/a,Johnston:2002/a} and the
magnitude of the dipole moment \cite{Johnston:2002/b}. Xu {\it et al.}
studied composite molecules made up of repulsive soft spheres, and found
isotropic and tilted crystalline phases \cite{Xu:2001/a}. More recently,
we have studied composite molecules made up of Lennard-Jones spheres -- so
called `composite Lennard-Jones molecules' (CLJMs) -- which exhibit
isotropic, nematic, tilted smectic, and tilted crystalline phases
\cite{PJC:2004/a}.

In this work the effects of molecular dipole moments and molecular
flexibility on the phase behavior of model bent-core molecules are studied
using computer simulations. The model (to be detailed in Section
\ref{sec:simulations}) consists of a rigid `V'-shaped core of soft spheres
with a point dipole moment oriented along the $C_{2}$ axis. Molecular
flexibility is included by the addition of short flexible tails to either
end. There is a substantial literature on the effects of these molecular
characteristics on linear molecules.  In hard-spherocylinder fluids, the
addition of longitudinal molecular dipoles is seen to destabilize the
nematic phase, and can even destabilize smectic phases if the dipoles are
displaced toward the ends of the molecules
\cite{McGrother:1996/b,McGrother:1998/a}; transverse dipoles also
destabilize the nematic phase with respect to the smectic A
\cite{Gil-Villegas:1997/a}. Gay-Berne ellipsoids with longitudinal point
dipoles show a stabilization of the nematic phase with respect to the
isotropic phase as the dipoles are moved from the centers of the molecules
to the ends \cite{Satoh:1996/a}, and can exhibit antiferroelectric smectic
phases with striped structures \cite{Berardi:1996/a}. Tilted polar smectic
phases have been reported in fluids of Gay-Berne molecules with transverse
dipole moments \cite{Gwozdz:1997/a}. As far as molecular flexibility is
concerned, the general consensus is that the introduction of flexible tail
groups destabilizes the nematic phases of hard spherocylinders
\cite{vanDuijneveldt:1997/a}, fused hard-sphere chains
\cite{McBride:2001/a,McBride:2002/a}, Gay-Berne \cite{Wilson:1997/a}, and
soft-sphere chains \cite{Affrouard:1996/a}. Interestingly, the
simultaneous presence of flexible tails and molecular dipole moments can
lead to a stabilization of the nematic phase
\cite{vanDuijneveldt:2000/a,Fukunaga:2004/a}. With regard to bent-core
molecules, Johnston {\it et al.} have shown that the presence of
transverse molecular dipoles on Gay-Berne dimers stabilizes the smectic
phases at the expense of nematic phases, increases the tilt angle in
tilted smectic phases, and can induce long-range polar ordering
\cite{Johnston:2002/b}. In the current work we will show that for the
bent-core soft-sphere models considered, the additions of dipolar
interactions and flexible tails both destabilize the nematic phase, and
that the dipolar interactions reduce the degree of molecular tilt in
smectic phases.

This paper also reports our attempts to measure the flexoelectric
coefficients \cite{Meyer:1969/a,deGennes:1993/a} of model bent-core
molecules. There are relatively few accounts of such measurements in the
literature. Experimentally, the determination of these quantities is
highly non-trivial \cite{Murthy:1993/a,Petrov:2001/a}, mainly due to the
fact that the flexoelectric coefficients are not measured directly, but
rather in linear combinations or as ratios involving elastic constants. In
simulations, the flexoelectric coefficients of pear-shaped Gay-Berne
ellipsoid/Lennard-Jones sphere molecules have been measured directly using
expressions involving the direct correlation function
\cite{Stelzer:1998/a}. The coefficients for a similar model were studied
in simulations using fluctuation expressions \cite{Billeter:2000/a}. The
compound 5CB has been studied using a parameterized dipolar Gay-Berne
model and the Percus-Yevick closure of the Ornstein-Zernike equation
\cite{Zakharov:2001/a,Zakharov:2002/a}. The flexoelectric coefficients
were computed using the direct correlation function route, and the results
compared moderately well with experiment \cite{Murthy:1993/a}. A very
recent simulation study of PCH5 -- using a fully atomistic molecular model
-- employed fluctuation formulae which yielded results in good agreement
with experiment \cite{Cheung:2004/a}. In the present work we critically
assess the reliability of the fluctuation route in the context of our
model systems, and show that the bend flexoelectric coefficients for
non-polar molecules can be comparable to those measured in experiments,
reflecting the significant role of molecular packing in dense liquids.

This paper is organized as follows. In Section \ref{sec:simulations} the
molecular model to be studied is fully defined, and the required
simulation methods are described. Simulation results for rigid linear
molecules are presented in Section \ref{sec:cssm00}, and those for rigid
bent-core molecules in Sections \ref{sec:cssm20} and \ref{sec:cssm40}. The
effects of molecular flexibility are considered in Section
\ref{sec:cssmt}, and flexoelectricity is discussed in Section
\ref{sec:flexo}. Section \ref{sec:conclusions} concludes the paper.

\section{Molecular model and simulation methods}
\label{sec:simulations}

\begin{figure}[!tbp]
\centering   
\includegraphics[scale=0.30]{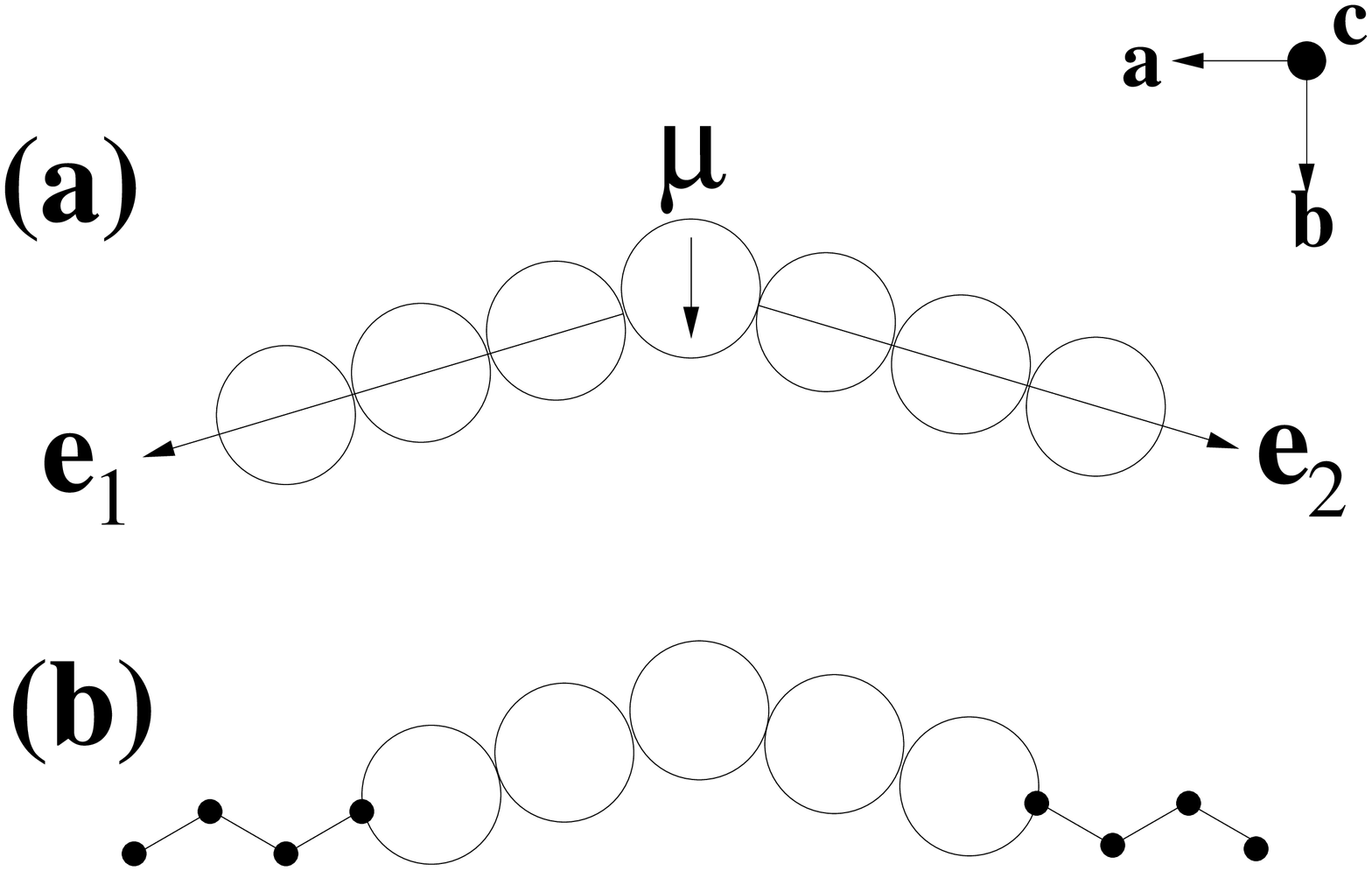}
\caption{\label{fig:models} The molecular models studied in this work: (a)
the CSSM model; (b) the CSSMT model. Also shown are the molecular axes,
${\bf a}$, ${\bf b}$, and ${\bf c}$.}
\end{figure}

The molecular models considered in this work are shown schematically in
Fig.~\ref{fig:models}. The most basic model [Fig.~\ref{fig:models}(a)]
consists of a rigid core of seven soft spheres arranged in a `V' shape
with external bond angle $\gamma$ defined such that ${\bf e}_{1}\cdot{\bf
e}_{2}=\cos{(180^\circ-\gamma)}$, where ${\bf e}_{1}$ and ${\bf e}_{2}$
are unit vectors pointing along the two `arms' of the molecule. The
sphere-sphere interaction potential is taken to be the repulsive part of
the Lennard-Jones (12,6), i.e.,
\begin{equation}
u_{ss}(r) = 4\epsilon\left(\frac{\sigma}{r}\right)^{12},
\label{eqn:uss}
\end{equation}
where $r$ is the sphere-sphere separation, $\epsilon$ is an energy
parameter, and $\sigma$ is the sphere `diameter'. In this work the
intramolecular sphere-sphere bond length is set equal to $1\sigma$. To
identify the molecular axes that will be aligned in orientationally
ordered phases, we define three unit vectors associated with the rigid
cores of the molecules, as illustrated in Fig.~\ref{fig:models}. These
vectors are given by ${\bf a}=({\bf e}_{1}-{\bf e}_{2})/|{\bf e}_{1}-{\bf
e}_{2}|$, ${\bf b}=({\bf e}_{1}+{\bf e}_{2})/|{\bf e}_{1}+{\bf e}_{2}|$
and ${\bf c}={\bf a}\wedge{\bf b}$.

Many real bent-core molecules possess a transverse electric dipole moment
aligned along the $C_{2}$ molecular symmetry axis. To represent
dipole-dipole interactions we have considered the model represented in
Fig.~\ref{fig:models}(a), where a point dipole moment is placed at the
center of the apical sphere along the $C_{2}$ symmetry axis (${\bf b}$).  
The dipole-dipole interaction is,
\begin{equation}
u_{dd}({\bf r},{\bm\mu}_{1},{\bm\mu}_{2}) = 
  \frac{{\bm\mu}_{1}\cdot{\bm\mu}_{2}}{r^{3}}
- \frac{3({\bm\mu}_{1}\cdot{\bf r})
         ({\bm\mu}_{2}\cdot{\bf r})}{r^{5}},
\label{eqn:udd}
\end{equation}
where ${\bf r}$ is the pair separation vector, $r=|{\bf r}|$, and
${\bm\mu}_{i}=\mu{\bf b}_{i}$ is the dipole vector on particle $i$. For
brevity, we shall refer to these molecules as `composite soft-sphere
molecules' (CSSMs).

Common additional features of real bent-core molecules include alkyl or
alkoxy chains (typically 3-12 carbons in length) attached to both ends of
each molecule. To represent these tails we consider the addition of four
extra spheres at each end of the model molecules, as represented in
Fig.~\ref{fig:models}(b). The resulting tail segments are allowed to pivot
under the following constraints: the tail bond length is equal to
$0.6\sigma$, which corresponds to the ratio of the carbon-carbon bond
length to the diameter of the aromatic ring in real bent-core liquid
crystals; each tail segment is oriented at the tetrahedral angle,
$\cos^{-1}{(-\mbox{$\frac{1}{3}$})}=109.47^\circ$, with respect to its
neighboring segments, mimicking the bond angles in a simple hydrocarbon
tail. Clearly this extension will cause a considerable increase in the
number of interactions to be evaluated, so to make the simulations
tractable we have made the following simplifications. Firstly, we have
reduced the size of the rigid bent-core to five spheres. (With the
addition of the tail segments, this actually makes the effective
elongations for all of the models considered more comparable.) Secondly,
we neglect interactions between tails on different molecules. Within the
Lorentz-Berthelot mixing rules \cite{Allen:1987/a}, this corresponds to
setting the tail-sphere diameter to zero. Denoting the core and tail
spheres by `c' and `t', respectively, the interaction potential is,
\begin{equation}
u_{ss}^{ij}(r) = 4\epsilon\left(\frac{\sigma_{ij}}{r}\right)^{12},
\label{eqn:ussij}
\end{equation}
with $i,j=$ c or t, $\sigma_{cc}=\sigma$, $\sigma_{tt}=0$, $\sigma_{ct} =
\mbox{$\frac{1}{2}$}(\sigma_{cc}+\sigma_{tt}) =
\mbox{$\frac{1}{2}$}\sigma$, and for simplicity the energy parameter
$\epsilon$ is the same for all pairs. This is clearly a very crude
representation of molecular flexibility, but it has proven to be an
appropriate means of extending the range of applicability of simple
liquid-crystal models \cite{vanDuijneveldt:1997/a,vanDuijneveldt:2000/a}.
Even though this is a simple model, configurational-bias MC techniques are
required to simulate the system efficiently; these are summarized in
Section \ref{sec:mc}.  For brevity, we shall refer to these molecules as
`composite soft-sphere molecules with tails' (CSSMTs).

Reduced units for these systems are defined as follows: reduced molecular
density, $\rho^{*}=N_{m}\sigma^{3}/V$, where $N_{m}$ is the number of
molecules and $V$ is the system volume; reduced temperature,
$T^{*}=k_{B}T/\epsilon$; reduced pressure, $p^{*}=p\sigma^{3}/\epsilon$;
reduced dipole moment, $\mu^{*}=\sqrt{\mu^{2}/\epsilon\sigma^{3}}$.

\subsection{Monte Carlo}
\label{sec:mc}

The phase behavior of the model systems was investigated using
constant-pressure ($NPT$) and constant-volume ($NVT$) Metropolis MC
simulations \cite{Allen:1987/a}. In all of the simulations reported in
this work, the number of molecules was $N_{m}=400$, with initial
high-density crystalline configurations consisting of four layers of 100
molecules. The general approach was to equilibrate the system at low
temperature ($T^{*} \sim 1$) and high density ($\rho^{*} \sim 0.14$) using
$NVT$ simulations in a cuboidal simulation cell with dimensions $L_{x} =
L_{y} \neq L_{z}$ and volume $V=L_{x}L_{y}L_{z}$, and then switching over
to $NPT$ simulations at a fixed pressure of $p^{*}=4$; earlier work
indicated that all of the expected liquid-crystalline phases could be
stabilized in a CLJM system at this pressure \cite{PJC:2004/a}.  
Simulations along the isobar were carried out at progressively higher
temperatures in order to locate phase transitions between solid, smectic,
nematic and isotropic phases; transitions were identified by measuring the
equation of state (density as a function of temperature) and relevant
order parameters (detailed below). Cooling runs were carried out using
cuboidal and/or cubic simulation cells to confirm the existence and nature
of the transitions. As explained in Ref.~\onlinecite{PJC:2004/a}, spot
checks on the stress tensor showed that the cuboidal/cubic cells did not
mechanically destabilize liquid-crystalline phases.

For the CSSM system, straightforward simulation techniques were employed
as detailed in earlier work \cite{PJC:2004/a}; single-particle translation
and rotation moves, and volume moves (in $\ln V$), were generated with
respective maximum displacement parameters to achieve $\sim 50\%$
acceptance rates. The long-range dipolar interactions were handled using
Ewald summations with conducting (`tin-foil') boundary conditions
\cite{Allen:1987/a}.

For the CSSMT system the CBMC technique was implemented and optimized as
described in Refs.~\onlinecite{Siepmann:1991/a,Frenkel:2002/a}. Tail
conformations were sampled by generating 5 trial orientations per segment
per MC move. Translational and rotational displacement parameters were
adjusted to give an acceptance ratio of $10\%$; in $NPT$ simulations the
volume moves were adjusted to give a $50\%$ acceptance ratio. The
computational effort required to simulate this system was considerable. To
carry out a MC sweep consisting of one attempted translation and rotation
per molecule, and one volume move, took approximately 4 seconds on a
$2.2~{\rm GHz}$ Intel Xeon processor; to achieve equilibration at each
state point required at least $10^{5}$ MC sweeps

To monitor orientational order, the order tensors ${\bf Q}_{\alpha\alpha}
= \mbox{$\frac{1}{2}$} \sum_{i=1}^{N} (3{\bm\alpha}{\bm\alpha}-{\bf 1})$
for each of the molecular axes ${\bm\alpha}={\bf a}$, ${\bf b}$, and ${\bf
c}$ were diagonalized yielding the eigenvalues $\lambda_{\alpha}^{-} <
\lambda_{\alpha}^{0} < \lambda_{\alpha}^{+}$, and the corresponding
orthonormal eigenvectors, ${\bf n}_{\alpha}^{-}$, ${\bf n}_{\alpha}^{0}$,
and ${\bf n}_{\alpha}^{+}$ \cite{Zannoni:1979/a}. The molecular $x$, $y$,
and $z$ axes were then assigned in order of increasing
$\lambda_{\alpha}^{+}$, and the laboratory axes, ${\bf X}$, ${\bf Y}$, and
${\bf Z}$, were identified with the corresponding directors. In practice
this almost invariably meant that the molecular $z$ axis was the `long'
axis ${\bf a}$ and the $x$ and $y$ axes were ${\bf b}$ and ${\bf c}$, and
that the laboratory $z$ axis was ${\bf n}_{a}^{+}$. The usual nematic and
biaxial order parameters -- $S$ and $Q_{22}^{2}$ - are then given by,
\begin{eqnarray} 
         S &=& {\bf Z}\cdot{\bf Q}_{zz}\cdot{\bf Z},\label{eqn:Q002ot}\\
Q_{22}^{2} &=& \mbox{$\frac{1}{3}$}(
               {\bf X}\cdot{\bf Q}_{xx}\cdot{\bf X} 
            +  {\bf Y}\cdot{\bf Q}_{yy}\cdot{\bf Y} \nonumber \\
           &-& {\bf X}\cdot{\bf Q}_{yy}\cdot{\bf X} 
            -  {\bf Y}\cdot{\bf Q}_{xx}\cdot{\bf Y}).
               \label{eqn:orderparameters}
\end{eqnarray}
In a perfect uniaxial nematic phase, $S=1$ and $Q_{22}^{2}=0$, whereas in
a perfect biaxial phase, $S=1$ and $Q_{22}^{2}=1$. In practice no biaxial
ordering was detected in any of the simulations, and so we will not report
the numerical values of $Q_{22}^{2}$ (which are less than $\sim 0.1$).  
Some additional measured observables include the polarization, ${\bf
P}=\mu\sum_{i}{\bf b}_{i}$, and the intermolecular torque-density tensor,
${\bm\Pi}$. In particular, these quantities are required for the
calculation of flexoelectric coefficients, full details of which will be
given in Section \ref{sec:flexo}.

\section{Results}
\label{sec:results}

\subsection{CSSMs with $\gamma=0^\circ$}
\label{sec:cssm00}

\begin{figure}[!tbp]
\centering
\includegraphics[scale=0.30,angle=270]{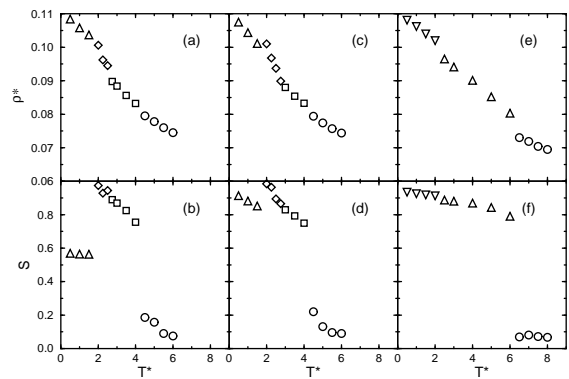}
\caption{\label{fig:cssmg00} Equations of state [(a),(c),(e)] and order
parameters [(b),(d),(f)] for CSSM systems with $\gamma=0^\circ$ along an
isobar with $p^{*}=4$: (a),(b) $\mu^{*}=0$; (c),(d) $\mu^{*}=1$; (e)m(f)
$\mu^{*}=2$. The symbols denote different phases: solid (down triangles);
tilted smectic B (up triangles); smectic A (diamonds); nematic (squares);
isotropic (circles).}
\end{figure}

\begin{figure}[!tbp]
\centering
\includegraphics[scale=1]{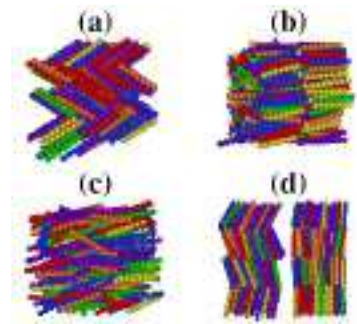} 
\caption{\label{fig:cssmg00snaps} Simulation snapshots of the CSSM system
with $\gamma=0^\circ$ (linear molecules) along an isobar with $p^{*}=4$:
(a) tilted smectic-B ($\mu^{*}=0$, $T^{*}=2$); (b) smectic-A ($\mu^{*}=0$,
$T^{*}=2.5$); (c) nematic ($\mu^{*}=0$, $T^{*}=4$); (d) tilted smectic-B 
($\mu^{*}=1$, $T^{*}=2$).}
\end{figure}

The equation of state and order parameters for apolar ($\mu^{*}=0$) linear
CSSMs along an isobar with $p^{*}=4$ are shown in
Figs.~\ref{fig:cssmg00}(a) and \ref{fig:cssmg00}(b), respectively. In
order of increasing temperature we find a high-density tilted phase ($0.5
\leq T^{*}\leq 1.5$), a smectic A ($2.0 \leq T^{*}\leq 2.5$), a uniaxial
nematic ($2.75 \leq T^{*}\leq 4$), and ultimately the isotropic phase
($T^{*} \geq 4.5$). Simulation snapshots are shown in
Fig.~\ref{fig:cssmg00snaps}. In the tilted phase, the molecules are
arranged in layers tilted by about $60^\circ$ with respect to each other,
as shown in Fig.~\ref{fig:cssmg00snaps}(a). This `herringbone' structure
clearly allows close-packing of the constituent spheres. It is difficult
to resolve the molecules in to layers unambiguously, but it is quite clear
from examining simulation snapshots that there is no long-range
crystalline order. In the absence of such order we therefore classify this
phase as tilted smectic B, if only to indicate that it is not crystalline
\cite{deGennes:1993/a}. We found no stable crystalline phase at $T^{*}
\geq 0.5$. Referring to Fig.~\ref{fig:cssmg00}(b), the jump in nematic
order parameter in the temperature range $1.5 < T^{*} < 2.0$ is due to the
transition from the tilted smectic-B phase to the untilted smectic-A
phase; the drop in the range $4 < T^{*} < 4.5$ signals the smectic
A-nematic transition. In all, the results for this system are in good
qualitative correspondence with those for a whole host of similar (linear)
molecular models, including soft-sphere chains \cite{Paolini:1993/a}, and
Lennard-Jones chains \cite{Galindo:2003/a}. The results are also
comparable to those presented in Ref.~\onlinecite{PJC:2004/a} for
seven-sphere CLJM fluids in which the sphere-sphere interaction is given
by $4\epsilon[(\sigma/r)^{12}-(\sigma/r)^{6}]$. Qualitatively, the CSSM
and CLJM systems are very similar, but in the latter case the phase
transitions are shifted to higher temperatures due to the attractive
component of the interaction potential.

With the addition of a small molecular dipole ($\mu^{*}=1$) we see little
qualitative difference in the equation of state at $p^{*}=4$, as shown in
Fig.~\ref{fig:cssmg00}(c). Despite the small change in the equation of
state, the gross structure of the tilted smectic-B phase is quite
different from that in the apolar system. The low-temperature smectic-B
phase is not so strongly tilted as in the apolar system, exhibiting a tilt
angle with respect to the layer normal of $\sim 20^\circ$. This is most
likely to allow dipoles on neighboring molecules to attain the low-energy
`nose-to-tail' conformation within the plane of the layer. The nematic
order parameter -- shown in Fig.~\ref{fig:cssmg00}(d) -- is relatively
high at temperatures $T^{*} \leq 1.5$ due to the reduced degree of tilt.
The tilted smectic B-smectic A and smectic A-nematic transitions are
signaled by changes in $S$ at $1.5 < T^{*} < 2$ and $4 < T^{*} < 4.5$,
respectively.

\begin{figure}[!tbp]
\centering
\includegraphics[scale=1]{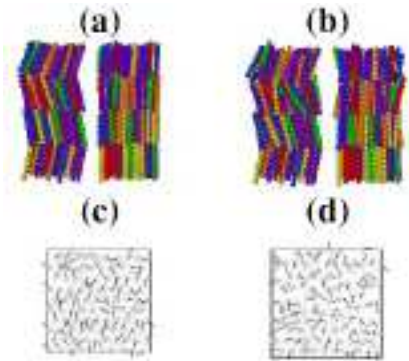}
\caption{\label{fig:cssmg00mu2snaps} Simulation snapshots of the CSSM
system with $\gamma=0^\circ$ and $\mu^{*}=2$: (a) crystalline phase
($T^{*}=2$); (b) tilted smectic-B phase ($T^{*}=4$); (c) a typical layer
in the crystalline phase ($T^{*}=2$); a typical layer in the tilted
smectic-B phase ($T^{*}=4$). In (c) and (d) the short black lines indicate
the orientations of the molecular dipole moments.}
\end{figure}

With a large dipole moment ($\mu^{*}=2$) and $p^{*}=4$ we see some
dramatic differences in the phase behavior: the nematic and smectic-A
phases are completely absent, and there is instead a distinct transition
between two high-density layered phases in the range $2 \leq T^{*} \leq
2.5$. Simulation snapshots at temperatures of $T^{*}=2$ and $T^{*}=4$ are
shown in Fig.~\ref{fig:cssmg00mu2snaps}. Examination of the layers
[Fig.~\ref{fig:cssmg00mu2snaps}(c) and \ref{fig:cssmg00mu2snaps}(d)] shows
that at $T^{*} \leq 2$ the system is in a crystalline phase, with
apparently long-range positional order within the layers. At $T^{*}=4$ the
in-layer ordering is qualitatively different, showing short-range
positional correlations and defects that destroy long-range positional
order. The equation of state and nematic order parameter are shown in
Figs.~\ref{fig:cssmg00}(e) and \ref{fig:cssmg00}(f). We assign the
branches in the equation of state as corresponding to crystalline ($0.5
\leq T^{*} \leq 2.0$), tilted smectic-B ($2.5 \leq T^{*} \leq 6.0$), and
isotropic ($T^{*} \geq 6.5$) phases.

To summarize, the addition of dipolar interactions to linear seven-sphere
molecules leads to a reduction in the degree of tilt in the
low-temperature smectic-B phase. With high dipole moments, the smectic-A
and nematic phases disappear, and a solid-smectic B transition is shifted
in to the range of temperatures considered in this work.

\subsection{CSSMs with $\gamma=20^\circ$}
\label{sec:cssm20}

\begin{figure}[!tbp]
\centering
\includegraphics[scale=0.30,angle=270]{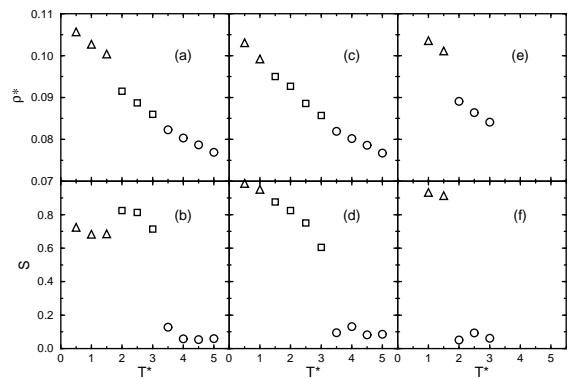}
\caption{\label{fig:cssmg2040} Equations of state [(a),(c),(e)] and order
parameters [(b),(d),(f)] for CSSM systems with $\gamma=20^\circ$ and
$\gamma=40^\circ$ along an isobar with $p^{*}=4$: (a),(b)
$\gamma=20^\circ$ and $\mu^{*}=0$; (c),(d) $\gamma=20^\circ$ and
$\mu^{*}=1$; (e),(f) $\gamma=40^\circ$ and $\mu^{*}=0$. The symbols denote
different phases: tilted smectic B (up triangles); nematic (squares);
isotropic (circles).}
\end{figure}

\begin{figure}[!tbp]
\centering
\includegraphics[scale=0.75]{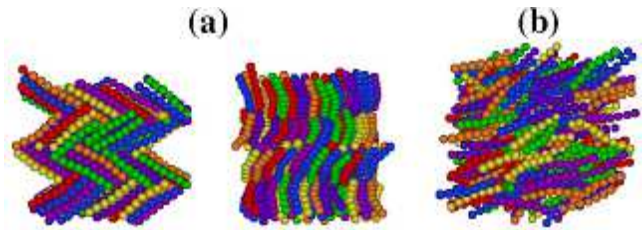}
\caption{\label{fig:cssmg20snaps} Simulation snapshots of the CSSM system
with $\gamma=20^\circ$ along an isobar with $p^{*}=4$: (a) two views of
the tilted smectic B phase at $T^{*}=1$; (b) nematic phase at $T^{*}=3$.}
\end{figure}

The equations of state and nematic order parameters for CSSM systems with
$\gamma=20^\circ$ along an isobar with $p^{*}=4$ are shown in
Fig.~\ref{fig:cssmg2040}. Results are shown for two dipole moments,
$\mu^{*}=0$ [Figs.~\ref{fig:cssmg2040}(a) and ~\ref{fig:cssmg2040}(b)] and
$\mu^{*}=1$ [Figs.~\ref{fig:cssmg2040}(c) and ~\ref{fig:cssmg2040}(d)].
Both the apolar and polar systems exhibit tilted smectic-B, nematic, and
isotropic phases; examples of the smectic and nematic phases in the
$\mu^{*}=0$ system are illustrated in Fig.~\ref{fig:cssmg20snaps}. In the
smectic-B phases it was observed that the degree of molecular tilt with
respect to the layer normal is far greater in the apolar case ($\sim
53^\circ$) than in the polar case ($<20^\circ$). The smectic B-nematic
phase transition appears to be more pronounced in the apolar system than
in the polar system, as evidenced by the associated features in the
equations of state and in the variations of the nematic order parameters.
Some simulations were attempted with $\mu^{*}=2$ but these suffered from
convergence problems; simulations with different initial configurations
failed to converge on to the same branch of the equation of state. It is
possible that this was due to the combination of the steric dipole
(molecular bend) and the `electric' dipole resulting in strong anisotropic
interactions and prohibitively slow convergence.

A comparison of Figs.~\ref{fig:cssmg00} and \ref{fig:cssmg2040} shows that
the presence of a modest molecular bend leads to the smectic-A phase being
destabilized. This same trend was observed in simulations of the CLJM
system \cite{PJC:2004/a}. The introduction of dipolar interactions to the
bent-core model then seems to stabilize the nematic phase slightly in
favor of the smectic B. The smectic-B phases themselves are tilted, but
the degree of tilt is reduced significantly upon the addition of dipolar
interactions. This perhaps provides a clue as to why dipolar interactions
apparently disfavor the smectic B; the bent-cores want to form a tilted
phase, but the dipolar interactions want an untilted phase as explained in
Section \ref{sec:cssm00}.

\subsection{CSSMs with $\gamma=40^\circ$}
\label{sec:cssm40}

The convergence problems encountered with the $\gamma=20^\circ$ system
were exacerbated by an increase of the bend angle to $\gamma=40^\circ$. In
this case it was only possible to achieve reliable results for the apolar
system. Results at $\mu^{*}=0$ and $p^{*}=4$ are shown in
Fig.~\ref{fig:cssmg2040}(e) and \ref{fig:cssmg2040}(f). The equation of
state and order parameters show only two branches, which correspond to
tilted smectic-B and isotropic phases. This is very similar to the
situation in the CLJM system with the same bend angle \cite{PJC:2004/a},
albeit with the CSSM system undergoing a phase transition at lower
temperature.

\subsection{CSSMTs}
\label{sec:cssmt}

\begin{figure}[!tbp]
\centering
\includegraphics[scale=1]{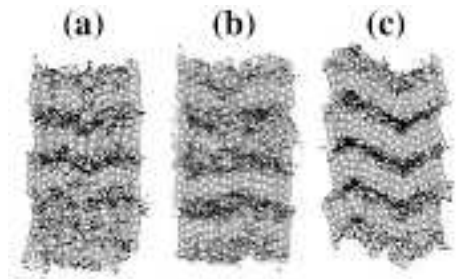}
\caption{\label{fig:cssmtsnaps} Simulation snapshots of CSSMT systems at
$T^{*}=1$ and $p^{*}=4$: (a) $\gamma=0^\circ$; (b) $\gamma=20^\circ$; (c) 
$\gamma=40^\circ$.}
\end{figure}

We performed $NPT$ simulations of CSSMT systems with bend angles of
$\gamma=0^\circ$, $20^\circ$, and $40^\circ$ along an isobar with
$p^{*}=4$. Due to the computational effort required for these simulations
we were not able to map out equations of state as comprehensive as those
for the CSSM systems. The simulation results are presented in Table
\ref{tab:cssmt}. The only phases observed in our simulations were smectic
and isotropic. In all cases the smectics were stable at $T^{*} \leq 1.5$,
and the smectic-isotropic transition occurred in the range $1.5 \leq T^{*}
\leq 2$. Some simulation snapshots of the smectic phases are shown in
Fig.~\ref{fig:cssmtsnaps}. In all cases the translational ordering within
the layers was of the smectic-B type, i.e., local hexagonal coordination.
The systems with $\gamma=0^\circ$ and $\gamma=20^\circ$ showed no
appreciable molecular tilt within the smectic layers, while the
$\gamma=40^\circ$ system showed an unusual `grain-boundary' structure
between clearly demarcated domains of untilted smectic. Larger-scale
simulations will be required to determine whether this is a signal of a
long-wavelength modulated structure. It is particularly striking that the
smectic-layer fluctuations are much larger than those in the CSSM (and
CLJM \cite{PJC:2004/a})  systems. Indeed, one criticism of the latter
models -- and other rigid-rod models -- is that the smectics are too well
ordered. Unsurprisingly, the introduction of molecular flexibility has
improved the correspondence between simulated smectic structures, and
those inferred from light-scattering experiments on common (flexible or
semiflexible)  mesogens \cite{deGennes:1993/a}.
\begin{table*}
\begin{center}
\caption{\label{tab:cssmt} Results from $NPT$ simulations of the CSSMT
system along an isobar with $p^{*}=4$. Digits in brackets denote the
estimated statistical uncertainty in the last figure.}
\begin{tabular}{dddddd}\hline\hline
\multicolumn{1}{c}{$\gamma$ / degrees} &
\multicolumn{1}{c}{$T^{*}$} &
\multicolumn{1}{c}{$\rho^{*}$} &
\multicolumn{1}{c}{$S$} &
\multicolumn{1}{c}{$l/\sigma$} &
\multicolumn{1}{c}{$\gamma_{\rm eff}$ / degrees} \\ \hline
 0 & 1.0 & 0.1098(4) & 0.918 & 5.48(1) & 48.0(3) \\
 0 & 1.5 & 0.1018(4) & 0.725 & 5.82(1) & 29.6(2) \\
 0 & 2.0 & 0.1067(7) & 0.091 & 5.40(3) & 31.8(3) \\
 0 & 2.5 & 0.1019(7) & 0.097 & 5.45(3) & 32.3(3) \\
 0 & 3.0 & 0.0981(7) & 0.063 & 5.44(2) & 32.5(3) \\
20 & 1.0 & 0.1128(2) & 0.865 & 5.41(1) & 50.9(4) \\
20 & 1.5 & 0.1101(2) & 0.787 & 5.51(1) & 43.3(2) \\
20 & 2.0 & 0.1057(7) & 0.200 & 5.35(3) & 37.0(5) \\
20 & 2.5 & 0.1010(3) & 0.062 & 5.34(3) & 37.4(4) \\
20 & 3.0 & 0.0976(7) & 0.054 & 5.34(2) & 37.6(4) \\
40 & 1.0 & 0.1161(3) & 0.879 & 4.98(1) & 65.2(4) \\
40 & 1.5 & 0.1084(1) & 0.833 & 4.93(1) & 60.7(2) \\
40 & 2.0 & 0.1048(6) & 0.067 & 5.08(3) & 49.5(3) \\
40 & 2.5 & 0.1008(6) & 0.081 & 5.10(2) & 49.2(6) \\
40 & 3.0 & 0.0974(7) & 0.057 & 5.11(2) & 49.4(5) \\
\hline\hline
\end{tabular}
\end{center}
\end{table*}

\begin{figure}[!tbp]
\centering
\includegraphics[scale=0.30,angle=270]{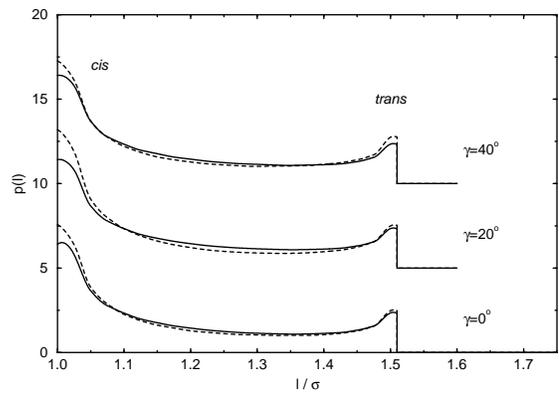}
\caption{\label{fig:cssmttorsion} Tail-length distribution functions for 
the terminal four-center tails in CSSMT systems along an isobar with
$p^{*}=1.0$: (bottom) $\gamma=0^{\circ}$; (middle) $\gamma=20^{\circ}$;
(top) $\gamma=40^{\circ}$. The curves for each bond angle are displaced by
10 units along the ordinate for clarity. In each case the solid lines
correspond to $T^{*}=1$ and the dashed lines to $T^{*}=3$. With bond
lengths of $0.6\sigma$ the pure {\it cis} conformation corresponds to     
$l=(5/3)\times 0.6\sigma=\sigma$, and the pure {\it trans} conformation to
$l=\sqrt{19/3}\times 0.6\sigma \simeq 1.51\sigma$.}
\end{figure}

The conformations of the flexible tail groups were investigated using some
simple measures. The extension of each tail was identified with the
distance, $l$, between the first and fourth joint (the black spheres in
Fig.~\ref{fig:models}). The probability density function, $p(l)$, is shown
in Fig.~\ref{fig:cssmttorsion} for all of the CSSMT systems at $T^{*}=1$
(smectic B) and $T^{*}=3$ (isotropic). Each function shows peaks at
$l/\sigma=1$ and $l/\sigma\simeq 1.5$. For a perfect {\it cis}
conformation the tail extension is $(5/3)$ times the bond length, while
for the {\it trans} conformation it is $\sqrt{19/3}$ times the bond
length. With the bond length being $0.6\sigma$, these distances correspond
to $l/\sigma=1$ and $l/\sigma\simeq 1.51$, respectively. Firstly, the {\it
cis} conformation is clearly the more favorable, presumably because the
molecules strive to attain the shortest effective elongation to minimize
excluded-volume interactions. Interestingly, for each system the {\it cis}
conformation appears slightly more favorable in the isotropic phase than
in the smectic phase. This may be due to the opportunity for
interdigitation of the tails with the cores in neighboring smectic layers,
which would explain the accompanying increase in the occurrence of the
{\it trans} conformation.

The effective molecular bend angle was determined by calculating the dot
product of the two unit vectors linking the apical sphere to the terminal
tail units, ${\bf u}_{1}$ and ${\bf u}_{2}$; the required angle is then
$\gamma_{\rm eff}=\cos^{-1}{(-{\bf u}_{1}\cdot{\bf u}_{2})}$. Values of
$\gamma_{\rm eff}$ are reported in Table \ref{tab:cssmt}. For each system
$\gamma_{\rm eff}>\gamma$ which shows that the tails must curl up in such
a way as to make the molecule more banana shaped. In the smectic phases
this can be observed directly in Fig.~\ref{fig:cssmtsnaps}, where the
tails prefer to be oriented in the plane of the smectic layers, rather
than pointing straight down toward the neighboring layers. This idea is
confirmed by the fact that the effective molecular bend is more pronounced
in the smectic phase than in the isotropic phase.

In summary, the addition of molecular flexibility results in the
disappearance of the nematic phase, and in the case of the linear
molecules, the smectic-A phase as well. This is in good qualitative
agreement with the trends observed in a variety of other liquid-crystal
models \cite{vanDuijneveldt:1997/a,McBride:2001/a,McBride:2002/a,%
Wilson:1997/a,Affrouard:1996/a,Fukunaga:2004/a}. The smectic-layer
structures of the flexible model systems correspond more closely to those
in real smectic liquid crystals.

\subsection{Flexoelectric coefficients}
\label{sec:flexo}

In 1969 Meyer predicted the existence of what is now known as the {\em
flexoelectric effect} in nematic liquid crystals, in which long-wavelength
distortions of the local molecular alignment (director field, ${\bf n}$)
give rise to a bulk polarization, ${\bf P}$ \cite{Meyer:1969/a}. The
textbook explanation of the effect is that if the director field possesses
curvature then an asymmetric molecular shape can dictate a favorable
local packing arrangement which, in the presence of molecular dipoles, may
give rise to a polarization \cite{deGennes:1993/a}.  With a splay
deformation ($\nabla\cdot{\bf n} \neq 0$) wedge-shaped molecules with
longitudinal dipole moments pack most efficiently when the dipoles are
aligned. With a bend deformation ($|\nabla\wedge{\bf n}| \neq 0$)
banana-shaped molecules with transverse dipole moments are arranged
preferentially to give a net polarization. General symmetry arguments lead
to the following relationship between the polarization density, ${\bf
p}=V^{-1}{\bf P}$ (units $\text{C}~\text{m}^{-2}$), and the lowest order
deformations of the director field,
\begin{equation}   
{\bf p} = e_{1}(\nabla\cdot{\bf n}){\bf n}
        + e_{3}(\nabla\wedge{\bf n})\wedge{\bf n}.
\label{eqn:p}
\end{equation}
where $e_{1}$ and $e_{3}$ are the splay and bend flexoelectric
coefficients, respectively, with units of $\text{C}~\text{m}^{-1}$.  
Historically there is some ambiguity in the sign of $e_{3}$; to be clear,
throughout this work we employ the convention used by Meyer in his
original study \cite{Meyer:1969/a}, Nemtsov and Osipov in their analysis
of flexoelectricity in the context of linear-response theory
\cite{Nemtsov:1986/a}, and de Gennes and Prost in their canonical text
\cite{deGennes:1993/a}. Allen and Masters have supplied a comprehensive
account of various simulation methods for measuring the flexoelectric
coefficients \cite{Allen:2001/a}. Following the sign conventions in
Ref.~\onlinecite{Allen:2001/a} we have calculated $e_{1}$ and $e_{3}$
using the relationships,
\begin{eqnarray} 
e_{1} &=& \mbox{$\frac{1}{2}$}\beta V^{-1}
           \left(
            \langle P_{z}\Pi_{xy} \rangle -
            \langle P_{z}\Pi_{yx} \rangle
           \right),
          \label{eqn:e1} \\
e_{3} &=& \mbox{$\frac{1}{2}$}\beta V^{-1}
           \left(
            \langle P_{y}\Pi_{zx} \rangle -
            \langle P_{x}\Pi_{zy} \rangle
           \right).
          \label{eqn:e3}
\end{eqnarray}
where ${\bm\Pi}=-\sum_{i<j}{\bf r}_{ij}{\bm\tau}_{ij}$ is the
orientational stress density tensor, ${\bf r}_{ij}={\bf r}_{i}-{\bf
r}_{j}$ is the intermolecular separation vector, and ${\bm\tau}_{ij}$ is
the torque on molecule $i$ due to molecule $j$. ${\bm\tau}_{ij}$ was
calculated as a sum of moments of the sphere-sphere interactions about the
apical sphere, and all vectors and tensors were calculated in a frame in
which the laboratory $z$ axis coincides with the nematic director, ${\bf
n}_{a}^{+}$. It is easy to show that the combinations
$(P_{z}\Pi_{xy}-P_{z}\Pi_{yx})$ and $(P_{y}\Pi_{zx}-P_{x}\Pi_{zy})$ are
invariant with respect to a rotation of the $x$ and $y$ axes about ${\bf
n}$, and so the assignments of the $x$ and $y$ axes are arbitrary. It
should be noted that we have only calculated the flexoelectric
coefficients for non-polar systems. The `steric dipole' is still parallel
to ${\bf b}$ in Fig.~\ref{fig:models}, but there are no {\em
electrostatic} dipole-dipole interactions. The polarization is given by
${\bf P}=\mu\sum_{i=1}^{N_{m}}{\bf b}_{i}$, and carries the trivial factor
of $\mu$ by virtue of there being no electrostatic interactions.

\begin{figure}[!tbp]
\centering
\includegraphics[scale=0.30,angle=270]{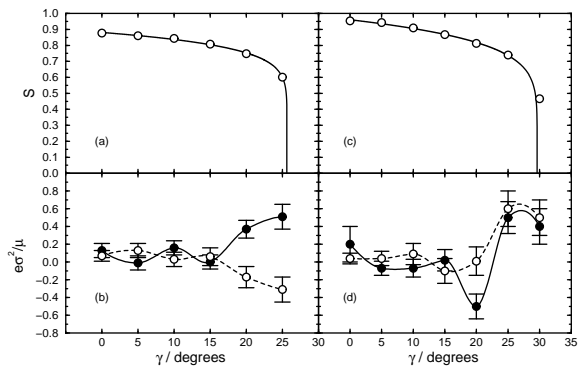}
\caption{\label{fig:flexo} Nematic order parameters [(a) and (c)] and     
flexoelectric coefficients [(b)  and (d)] for the CSSM system at
$\rho^{*}=0.0888$ and $T^{*}=3$ [(a) and (b)] and the CLJM system at
$\rho^{*}=0.1155$ and $T^{*}=6.5$ [(c) and (d)]. In (b) and (d) the  
flexoelectric coefficients are those corresponding to splay deformations
(filled symbols) and bend deformations (open symbols). The lines are 
spline fits to guide the eye.}
\end{figure}

The flexoelectric coefficients have been calculated in the nematic phases
of non-polar CSSM and CLJM \cite{PJC:2004/a} systems as a function of the
molecular bend angle, $\gamma$. For the purposes of comparison, the CSSM
system has been studied at a fixed density and temperature for which the
nematic phase is stable at several values of $\gamma$. An examination of
Figs.~\ref{fig:cssmg00}(a) and \ref{fig:cssmg2040}(a) shows that in the
range $0^\circ \leq \gamma \leq 20^\circ$, the nematic phase is stable at
temperatures and densities in the regions of $T^{*} \sim 3$ and
$\rho^{*}\sim 0.085$, respectively. A particular state point from the
$\gamma=0^\circ$ system was selected arbitrarily for all of the
simulations, this being $T^{*}=3$ and $\rho^{*}=0.0888$. Canonical ($NVT$)
simulations were used to equilibrate nematic phases for systems with bend
angles in the range $0^\circ \leq \gamma \leq 25^\circ$. The nematic order
parameter, $S$, is shown as a function of $\gamma$ in
Fig.~\ref{fig:flexo}(a). We found that $S$ could be fitted with a power
law,
\begin{equation}
S(\gamma) = S(0)\left( 1 - \frac{\gamma}{\gamma_{c}} \right)^\alpha,
\label{eqn:sgamma}
\end{equation}
where $S(0)$ is the order parameter for linear molecules, $\gamma_{c}$ is
a critical bend angle above which the nematic phase is no longer
thermodynamically stable, and $\alpha$ is a specific exponent. The fit is
shown in Fig.~\ref{fig:flexo}(a); the fit parameters were $S(0)=0.881(4)$,
$\gamma_{c}=25.6(2)^\circ$, and $\alpha=0.104(7)$. A similar procedure was
carried out for the CLJM system studied in Ref.~\onlinecite{PJC:2004/a}.
Nematic phases were simulated at $T^{*}=6.5$ and $\rho^{*}=0.1155$ for
systems with $0^\circ \leq \gamma \leq 30^\circ$. The nematic order
parameter is shown as a function of $\gamma$ in Fig.~\ref{fig:flexo}(c). A
power-law fit yielded the parameters $S(0)=0.961(4)$,
$\gamma_{c}=29.6(1)^\circ$, and $\alpha=0.143(4)$; the fit is shown in
Fig.~\ref{fig:flexo}(c).

Equations (\ref{eqn:e1}) and (\ref{eqn:e3}) were evaluated using results
from immense $NVT$-MC simulations. In general we carried out runs
consisting of $\sim 4 \times 10^{6}$ attempted MC translations and
rotations per molecule. In all cases the components of ${\bf P}$
fluctuated about zero, but long runs were required to ensure that the
average polarization, $\langle {\bf P} \rangle$, was almost zero. Results
for the reduced flexoelectric coefficients $e_{1}^{*}=e_{1}\sigma^{2}/\mu$
and $e_{3}^{*}=e_{3}\sigma^{2}/\mu$ in the CSSM and CLJM systems are
presented in Table \ref{tab:flexo} and Fig.~\ref{fig:flexo}. Also included
in Table \ref{tab:flexo} are the components of the average polarization,
${\bf P}$, and estimated uncertainties. In both the CSSM and CLJM systems,
the measured flexoelectric coefficients are small when the molecular bend
angles are less than about $20^\circ$. For more pronounced bend angles,
both the splay and the bend coefficients deviate significantly from zero.  
The results for $e_{3}^{*}$ are encouraging, but there is some concern
over the measured values of $e_{1}^{*}$; naively we would expect that the
splay coefficients should be small for banana-shaped molecules with
transverse dipoles. Table \ref{tab:flexo} yields a valuable clue; the
measured flexoelectric coefficients deviate significantly from zero in
those simulations where there is a more pronounced average polarization.
Therefore, it may be that the simulations are still not long enough to
ensure a reliable evaluation of the fluctuation formulae. It is perhaps
worth pointing out that during the course of the atomistic simulations
performed by Cheung {\it et al.} there was a `small net polarization'
\cite{Cheung:2004/a}: the magnitudes of the average polarizations ($P \sim
10^{-28}~\text{C}~\text{m}^{-1}$) and the average molecular dipole
moments ($\mu\sim 10^{-29}~\text{C}~\text{m}^{-1}$)  give reduced
polarizations $P/\mu \sim 10$, which are large compared to the average
polarizations reported in Table \ref{tab:flexo}. It is possible that
neither the estimates of Cheung {\it et al.} nor the current estimates are
particularly reliable. We attempted to make {\it ad hoc} corrections to
the fluctuation formulae in Eqns.~(\ref{eqn:e1}) and (\ref{eqn:e3}) by
evaluating terms like $\langle ( P_{\alpha} - \langle P_{\alpha} \rangle )  
( \Pi_{\beta\gamma} - \langle \Pi_{\beta\gamma} \rangle ) \rangle$, but
these resulted in insignificant changes to the values of $e_{1}$ and
$e_{3}$.
\begin{table*}
\begin{center}
\caption{\label{tab:flexo} Results from $NVT$ simulations of CSSM and CLJM
systems in the nematic phase at reduced density $\rho^{*}$ and reduced
temperature $T^{*}$. $\gamma$ is the molecular bend angle, $S$ is the
nematic order parameter, $e_{1}^{*}=e_{1}\sigma^{2}/\mu$ and
$e_{3}^{*}=\sigma^{2}/\mu$ are the reduced splay and bend flexoelectric
coefficients, respectively, and $P_{\alpha}$ is the average of the
$\alpha$ component of the system polarization. Digits in brackets denote
the estimated statistical uncertainty in the last figure based on two
standard deviations.}
\begin{tabular}{ddddddd}\hline\hline
\multicolumn{1}{c}{$\gamma$ / degrees} &
\multicolumn{1}{c}{$S$} &
\multicolumn{1}{c}{$e_{1}^{*}$} &
\multicolumn{1}{c}{$e_{3}^{*}$} &
\multicolumn{1}{c}{$P_{x}/\mu$} &
\multicolumn{1}{c}{$P_{y}/\mu$} &
\multicolumn{1}{c}{$P_{z}/\mu$} \\ \hline
\multicolumn{7}{c}{CSSM, $T^{*}=3$, $\rho^{*}=0.0888$} \\
 0 & 0.877 &  0.13(4) &  0.07(3) & -0.21(4) &  0.77(6) &  1.43(2) \\
 5 & 0.861 & -0.01(4) &  0.13(4) & -0.31(6) & -0.38(6) &  0.02(2) \\
10 & 0.844 &  0.16(4) &  0.03(4) &  0.27(6) & -0.38(6) & -0.10(2) \\
15 & 0.807 & -0.00(4) &  0.06(5) &  0.90(6) & -1.04(6) &  0.56(2) \\
20 & 0.749 &  0.37(5) & -0.17(6) &  0.09(6) & -0.55(6) &  0.67(4) \\
25 & 0.601 &  0.51(7) & -0.31(7) & -1.10(6) &  1.00(6) & -0.75(4) \\
\multicolumn{7}{c}{CLJM, $T^{*}=6.5$, $\rho^{*}=0.1155$} \\
 0 & 0.954 &  0.2(1)  &  0.04(3) & -0.13(6) & -0.78(6) & -0.13(4) \\
 5 & 0.943 & -0.07(4) &  0.04(4) &  0.09(6) & -0.42(6) &  0.13(2) \\
10 & 0.909 & -0.07(5) &  0.09(6) & -0.29(6) & -0.17(6) &  0.06(2) \\
15 & 0.867 &  0.02(6) & -0.10(7) & -0.18(6) & -1.06(6) &  0.10(2) \\
20 & 0.813 & -0.50(7) &  0.01(8) & -2.03(6) & -0.04(6) &  0.16(2) \\
25 & 0.739 &  0.50(9) &  0.6(1)  &  2.41(6) &  0.53(6) & -0.72(2) \\
30 & 0.466 &  0.40(1) &  0.5(1)  & -1.04(6) &  2.93(6) &  0.56(4) \\
\hline\hline
\end{tabular} 
\end{center}
\end{table*}

Notwithstanding the potential problems highlighted above, we can attempt
to make some useful comments on the measured values of $e_{3}$. At the
highest bend angles, $\gamma=20^\circ$-$30^\circ$, the magnitude of the
reduced bend flexoelectric coefficient is in the region of $0.5$. With
typical values of $\sigma \sim 0.5~\text{nm}$ and $\mu \sim
1~\text{Debye}$, this reduced value corresponds to a real bend
flexoelectric coefficient of $e_{3} \sim 7~\text{pC}~\text{m}^{-1}$. This
is in good agreement with typical values of $e_{3}\sim
10~\text{pC}~\text{m}^{-1}$ measured in experiments
\cite{Murthy:1993/a,Petrov:2001/a}. It is therefore reasonable to suggest
that the steric or packing contributions to the flexoelectric effect are
significant. The roles of dipolar and quadrupolar electrostatic
interactions must surely be at least as significant, and these should be
studied in a systematic fashion. For now, though, we conclude that
short-range interactions between bent-core molecules are as important in
giving rise to flexoelectricity as they are in dictating the short-range
structure of dense atomic liquids \cite{Hansen:1986/a}. In drawing this
analogy perhaps we should not be too surprised by the current
observations. The reduced densities of spheres in the CSSM and CLJM
systems are equal to $7\rho^{*} ~ \sim 0.7$ at which packing effects and
short-range correlations are particularly pronounced (recall that the
triple-point density for the Lennard-Jones system is in the region of
$0.85$).
 
\section{Conclusions}
\label{sec:conclusions}

In this work the structure, phase behavior, and flexoelectricity of model
bent-core molecules have been studied using MC computer simulations. The
molecular bent core consists of a `V' shaped rigid array of soft spheres,
with a transverse point dipole moment aligned along the $C_{2}$ symmetry
axis.

In a linear seven-sphere non-polar model, isotropic, nematic, smectic-A,
and tilted smectic-B (herringbone) phases are observed. With an opening
angle of $160^\circ$ the smectic A is absent, while an opening angle of
$140^\circ$ gives rise to a direct tilted smectic B -- isotropic
transition. The effects of dipolar interactions were seen to depend on the
opening angle. In the linear-molecule systems these interactions appeared
to destabilize the nematic and smectic-A phases. In the bent-core systems,
dipolar interactions reduced the degree of molecular tilt in the tilted
smectic phases; this is a significant observation, particularly since
dipolar interactions have often been cited as the cause of spontaneous
chiral symmetry breaking in some bent-core liquid crystals
\cite{Madhusudana:2004/a}. Chirality arises in this case from the
correlation between molecular tilt within smectic layers and the
long-range ordering of smectic-layer polarizations.  Although spontaneous
(anti)ferroelectric order will be favored by long-range dipolar
interactions, the observation that these same interactions reduce
molecular tilt seems to suggest that there may be another explanation for
chirality in banana liquid crystals
\cite{Lansac:2003/a,Emelyanenko:2004/a,Earl:2005/a}.

Real bent-core liquid crystals often possess flexible tail groups at the
ends of the rigid bent core, and so CBMC simulations of a
flexible-rigid-flexible model were performed. Each molecule consisted of a
five-sphere rigid bent core, and a three-segment flexible tail attached to
each end. We could only find smectic and isotropic phases, but
significantly the smectic phases showed no spontaneous tilt. This is
probably due to the tails providing a lubricating barrier between the
smectic layers that serves to decorrelate the order within neighboring
layers. Hence, the addition of molecular flexibility is likely to mitigate
against the type of entropic `sawtooth' mechanisms that have been found to
stabilize antiferroelectric ordering in hard-particle bent-core models
\cite{Lansac:2003/a}. The addition of the flexible tails was also seen to
give rise to significant spatial fluctuations in the smectic layers. It is
worth pointing out that the smectic phases of rigid model molecules are
often far more ordered than real smectics (as evidenced by scattering
experiments \cite{deGennes:1993/a}). The introduction of molecular
flexibility therefore brings the model systems in to better correspondence
with experiment.

Finally, the flexoelectric properties of non-polar seven-sphere bent-core
molecules - with and without attractive interactions - have been studied
by calculating the splay and bend coefficients in the nematic phase using
fluctuation relations derived from linear-response theory
\cite{Nemtsov:1986/a,Allen:2001/a}.  An immense investment of
computational effort was required to obtain reasonable results via this
route, which serves to highlight how careful one must be in evaluating the
required formulae. Nonetheless, our results show that a significant
flexoelectric response can be measured for opening angles below about
$150^\circ$. With typical molecular dimensions and dipole moments, the
measured flexoelectric coefficients are in the region of
$10~\text{pC}~\text{m}^{-1}$ which is in excellent agreement with
experiment. The flexoelectric response of real bent-core liquid crystals
is often attributed largely to dipolar and quadrupolar interactions, but
our results show that the molecular shape is also significant. This
shouldn't be too much of a surprise, since most thermotropic nematics are,
after all, dense molecular liquids, and it is well known that the
structure and dynamics in such systems are dictated by short-range
repulsive interactions. We are in no way suggesting that electrostatic
interactions are insignificant, and we have not studied flexoelectricity
in dipolar or quadrupolar systems because of the computational effort
which will probably be required to obtain reliable results. A systematic
study of this point is required, and will hopefully be the subject of
future papers.

\begin{acknowledgments}
The provision of a studentship for AD and computing hardware by the
Engineering and Physical Sciences Research Council (UK) (GR/R45727/01) is
gratefully acknowledged. We are grateful to Professor M. P. Allen
(Warwick) for helpful comments.
\end{acknowledgments}


\end{document}